\documentclass[times,authoryear]{elsarticle}

\usepackage{jasr}
\usepackage{framed,multirow}

\usepackage{amssymb}
\usepackage{latexsym}
\usepackage{amsmath}

\usepackage[switch]{lineno}

\usepackage{url}
\usepackage{xcolor}
\definecolor{newcolor}{rgb}{.8,.349,.1}

\usepackage[citebordercolor=white]{hyperref}

\journal{ASR}

\begin{document}

\verso{Chakraborty and Seemala}

\begin{frontmatter}

\title{Interhemispheric differences in field-aligned currents, ground magnetic perturbations, and TEC during the geomagnetic storms of May and October 2024}

\author[1,2]{Sumanjit Chakraborty\corref{c-d54cc1eb1ca4}}
\ead{sumanjit11@gmail.com}\cortext[c-d54cc1eb1ca4]{Corresponding author.}
\author[1]{Gopi K. Seemala}
\ead{gopi.seemala@gmail.com}

\affiliation[1]{organization={Indian Institute of Geomagnetism}, 
                postcode={Navi Mumbai},
                country={India}
                }

\affiliation[2]{organization={Presently at the Institute of Astronomy Space and Earth Science}, 
                postcode={Kolkata},
                country={India}
                }

\received{xx}
\finalform{xx}
\accepted{xx}
\availableonline{xx}
\communicated{}

\begin{abstract}

This study investigates storm-to-storm variability and hemispheric differences in magnetosphere–ionosphere (MI) coupling during the extreme (G5) geomagnetic storm of May 10-11 and the severe (G4) storm of October 10-11, 2024. Global field-aligned current (FAC) patterns derived from the Active Magnetosphere and Planetary Dynamics Response Experiment (AMPERE), together with conjugate observations from ground-based magnetometers within the SuperMAG network and Global Positioning System (GPS)-derived total electron content (TEC), are analyzed to examine high-latitude electrodynamic and ionospheric responses in both hemispheres. The May event exhibits broad and relatively organized Region 1/Region 2 FAC systems encircling the polar caps across multiple local time sectors, accompanied by intervals of correspondence in conjugate magnetic perturbations and structured TEC enhancements with temporal offsets between hemispheres. In contrast, the October event shows more localized, asymmetric, and uneven FAC morphology with pronounced hemispheric and dawn–dusk asymmetries, together with greater divergence in conjugate magnetic responses and spatially heterogeneous TEC variability. These differences are consistent with enhanced mesoscale variability and asymmetric current closure under storm-time conditions. Overall, the results highlight that even under similarly strong solar wind driving, the coupled MI system can exhibit substantially different spatial organization and interhemispheric coupling, reflecting the combined influence of FAC morphology, ionospheric conductance, and local electrodynamic conditions.

\end{abstract}


\end{frontmatter}


\section{Introduction}

The basis of solar-terrestrial relations and associated physics is the interaction between the omnipresent solar wind (a continuous flow of charged particles emitted from the Sun's corona) and Earth's Magnetosphere-Ionosphere (MI) system. This solar wind carries the frozen-in Interplanetary Magnetic Field (IMF). When $B_z$, the north-south component of this IMF, turns southward \citep{sc:01,sc:02}, energy from the solar wind and momentum are transferred (via magnetic reconnection) to the near-Earth space environment, triggering large-scale flows of plasma and currents in the MI system. This process (whose sources are the solar storms that release huge amounts of energy in the form of Coronal Mass Ejections (CMEs) and flares) triggers geomagnetic storms, which manifest as disturbances in the geomagnetic field. Strong levels of solar wind particle injection into the magnetosphere drive strong toroidal currents (the ring current) at distances of about 4-7 Earth radii \citep{Daglis_IA}.
Geomagnetic storms are classified as intense when the SYM-H (in units of nT), a high-resolution version of the Disturbance storm time (Dst) index related to ring current intensification, drops below -100 nT (see: \cite{sc:03,sc:04,sc:05,sc:06,Echer_E,Chakraborty_S,Chakraborty_S2} and references therein).

The high-latitude regions play a critical role in this coupling process because they provide the primary pathway for magnetospheric energy deposition into the ionosphere through the Field-aligned Currents (FACs), auroral precipitation, and enhanced ionospheric convection (see: \cite{CHAKRABORTY20262602,MISHRA2026101411} and references therein). As a result, the polar ionosphere exhibits strong variability during geomagnetic disturbances, including significant perturbations in ground magnetic fields and substantial modifications of the ionospheric plasma density. Such disturbances can produce large fluctuations in Total Electron Content (TEC), which affect satellite navigation and radio communication systems \citep{sc:17,sc:18,sc:19,sc:20,sc:21,sc:22}. Understanding the spatial structure and hemispheric characteristics of these disturbances remains an important topic in space physics, particularly during severe-to-extreme geomagnetic storms.

A growing number of recent studies have demonstrated that MI coupling processes are often asymmetric between the northern and southern hemispheres. These asymmetries arise from several factors, including seasonal variations in solar illumination and ionospheric conductivity, non-dipolar features of the geomagnetic field, and IMF $B_y$-related magnetospheric forcing \citep{sc:41,Tenfjord_P,sc:42}. Observational and statistical studies using the Active Magnetosphere and Planetary dynamics Response Experiment (AMPERE), ground magnetometer networks, and ionospheric measurements have shown that the strength, morphology, and location of FACs and equivalent ionospheric currents can differ significantly between conjugate regions during geomagnetic activity \citep{Coxon_JC,Tenfjord_P,Coxon_JC1,Workayehu_AB}. These hemispheric differences influence how magnetospheric forcing is transmitted to the ionosphere and how ground magnetic perturbations and ionospheric plasma structures develop under geomagnetic storm-time conditions.

The geomagnetic storms of May and October 2024 represent two of the strongest disturbances observed during the present solar cycle. The May 2024 event, often referred to as the Gannon storm, produced exceptionally intense geomagnetic activity and widespread auroral displays at unusually low latitudes. Several recent studies have investigated different aspects of this event. For example, \citep{STR} examined the ionospheric response and TEC disturbances associated with the storm, while \citep{dMC} analyzed the global impact of this event using ground magnetometer observations. Additionally, \citep{Zou} investigated the extreme auroral electrojet activity and magnetospheric drivers associated with the storm, highlighting the complex dynamics of the nightside current system during this period. These studies collectively demonstrate that the May 2024 storm involved highly dynamic MI coupling processes and significant global disturbances. Likewise, the October event \citep{Denny,Kleimenova,Paul,Tripathi_SC} produced severe geomagnetic activity and substantial disturbances in the high-latitude MI system, providing another opportunity to investigate storm-time electrodynamic responses under strong solar wind driving.

Despite these recent advances, simultaneous interhemispheric comparisons of FAC structures, ground magnetic perturbations, and ionospheric TEC responses during these extreme events remain limited. In particular, relatively few studies have examined how the large-scale current systems observed by satellite measurements relate to conjugate ground magnetic and ionospheric responses in both hemispheres during these storms. Such comparisons can provide valuable insight into how magnetospheric forcing is communicated through the coupled MI system and how local ionospheric conditions modulate the resulting electrodynamic response.

In this study, we investigate the MI response during the May and October 10–11, 2024, geomagnetic storms using a combination of interplanetary observations, AMPERE FAC maps, ground magnetometer measurements from conjugate Antarctic–Arctic station pairs, and Global Positioning System (GPS)-derived TEC data. This study is motivated by the fact that during geomagnetic storms, FACs transfer magnetospheric stress to the high-latitude ionosphere, where they close through horizontal ionospheric current systems that produce magnetic perturbations observed at ground stations. The associated electrodynamic forcing can also influence ionospheric plasma transport through enhanced convection and Joule heating, leading to variations in electron density and TEC. By comparing these multi-instrument observations between the two events, we examine differences in the hemispheric electrodynamic response and explore how the spatial morphology of FACs relates to ground magnetic perturbations and ionospheric plasma variability. Through this approach, the study provides an observational perspective on storm-to-storm variability and hemispheric differences in the high-latitude response of the MI system during severe-to-extreme geomagnetic activity.

The manuscript is structured as follows: Section 2 describes the database and methodology. Section 3 presents the detailed results of the two events of May and October 2024. Section 4 presents the discussion, and the study concludes with a summary in Section 5.

\section{Database and methodology}

For the present study, the high-resolution (1-minute) and openly available data of the IMF components $B_y$ and $B_z$, the solar wind velocity $V_{sw}$ and the density, the y-component of the interplanetary electric field $IEF_y$, and the SYM-H index were obtained from the OMNIWeb database of the Space Physics Data Facility of GSFC/NASA. We have also used the SML index from the SuperMAG network. SuperMAG is a global collaboration of national organizations and agencies operating more than 300 ground-based magnetometers. They provide magnetic field variations in a common coordinate system, with the same time resolution and a common baseline removal approach \citep{sc:53,sc:54}. 

Next, we have used data from the Indian Antarctic stations, Maitri and Bharati, which have been operational since 1989 and 2013, respectively. The National Centre for Polar and Ocean Research (NCPOR) of the Ministry of Earth Sciences (MoES), Government of India, manages these stations. Additionally, this study uses measurements from SuperMAG and International GNSS Service (IGS) receivers. The magnetic field measurements obtained are post-nighttime base-level-subtracted, and all data have a 1-minute time resolution. Table \ref{stations} displays the locations of ground magnetometers and GPS receivers. We have consistently used the standard convention of 3-letter uppercase codes for magnetometers and 4-letter lowercase codes for GPS stations throughout this work. The Antarctic and Arctic stations considered here are selected to represent approximately magnetic conjugate locations in the two hemispheres. The magnetic conjugate points are estimated using an International Geomagnetic Reference Field (IGRF)-based field-line tracing from the Antarctic stations to the opposite hemisphere. Exact magnetic conjugacy between ground stations in the two hemispheres is rare. Therefore, nearby high-latitude Arctic observational stations located close to these estimated conjugates are used for the analysis. For the GPS station mtri, a Leica 1200 receiver collected raw pseudorange and phase data from GPS observables. The data were converted into Receiver Independent Exchange (RINEX) format using the TEQC program \citep{sc:61}. These RINEX files, together with those from other IGS stations ('arht', 'eil3', and qaq1'), were read by the GPS TEC program \citep{sc:43} to calculate the vertical TEC used in this study. In these calculations, we assumed the ionospheric pierce point was at 350 km altitude and used a single shell mapping function \citep{Mannucci_A} to convert slant TEC to vertical TEC. The full algorithm, including bias removal, is detailed in \citep{sc:29,sc:64}. 

\begin{table*}
\centering
\caption{Geographic (Geog.) coordinates (lat$^\circ$,lon$^\circ$) of Antarctic stations, their IGRF-based conjugate points, and the selected near-conjugate Arctic observational stations together with their geographic coordinates}
\vspace{10pt}
\begin{tabular}{|l|l|l|}
\hline
Antarctic station: code, Geog. & Conjugate Geog. point & Near-conjugate station: code, Geog.\\
\hline
Bharati: BRT (-69.41,76.20)  & 68.90,20.80  & Sørøya: SOR (70.54,22.22)  \\ 
\hline
Maitri: mtri (-70.77,11.73)  & 62.10,311.41 & Qaqortoq: qaq1 (60.72,313.95) \\ 
\hline
McMurdo: arht (-77.83,166.66) & 62.80,215.12 & Fairbanks: eil3 (64.69,212.89) \\ 
\hline
Neumayer III: VNA (-70.70,351.70) & 66.60,22.73  & Kiruna: KIR (67.84,20.42)  \\ 
\hline
Vostok: VOS (-78.45,106.87) & 77.30,12.55 & Ny-Ålesund: NAL (78.92,11.95)  \\
\hline
\end{tabular}
\label{stations}
\end{table*}

Further, we used data from the AMPERE, which provides global measurements of ionospheric magnetic perturbations and FAC densities derived from magnetometers onboard the Iridium satellite constellation. The AMPERE dataset enables continuous monitoring of large-scale FAC systems and their temporal evolution at high latitudes. In this study, we used the processed horizontal magnetic perturbation ($B_n$) and radial current density ($J_{par}$) data products, together with the corresponding polar maps, obtained from the AMPERE Science Data Center. These data products are derived using established inversion techniques that map satellite magnetic perturbations to ionospheric current systems \citep{Waters_CL,Anderson_BJ}. Subsequent studies have further characterized the variability and response timescales of AMPERE-derived FAC systems under varying solar wind and IMF conditions \citep{Anderson_BJ1,Coxon_JC,Coxon_JC1,Waters_CL1}. The AMPERE observations provide a global view of geomagnetic storm-time current morphology and are widely used to investigate MI coupling processes.

Finally, the NCAR Whole Atmosphere Community Climate Model with thermosphere and ionosphere eXtension (WACCMX) simulations of TEC and hmF2 were used. The WACCMX is a general circulation model providing a fully coupled, self-consistent representation of atmospheric chemistry and dynamics. It computes three-dimensional temperature, composition, winds, and ionospheric structure from the surface up to about 700 km altitude. High-latitude ionospheric conditions and solar spectral irradiance drive this model. Its outputs include electron and ion temperatures and densities; Hall and Pedersen conductivities; meridional, zonal, and vertical ion drifts; neutral winds; and species compositions such as $O$, $O_2$, $NO$, $H$ \citep{sc:47,sc:48}.

\section{Results}

In this section, ionospheric responses to the G5 class (Extreme, according to NOAA Space Weather Scales) event during May 10-11 and a G4 (Severe) class during October 10-11, 2024, are presented in detail. The interplanetary and geomagnetic parameters are presented first, followed by the AMPERE maps of magnetic field perturbations and current densities. Next, we present observations of $\Delta H$ variations from the near-conjugate Antarctic-Arctic station pairs and conclude the section by showing the temporal variation of TEC along with AMPERE-derived magnetic field and current density perturbation time series.

\subsection{Extreme event of May 10-11, 2024}

The Gannon storm (in memory of space physicist \textit{Jennifer L. Gannon}) is a historic space weather event and one of the strongest since the March 1989 Qu\'ebec storm. The manifestations of the effects of this storm on the terrestrial environment were the aurora sightings from both hemispheres at much lower latitudes (around 30-35$^\circ$) than usual \citep{STR,Hayakawa_H}. The event was associated with a sequence of powerful solar eruptions originating from active region (AR) 3664 between May 8 and 10, which produced multiple interacting CMEs and complex interplanetary structures impacting the Earth \citep{Hayakawa_H,Zou,Gordiyenko}. Previous investigations have examined different aspects of this event, including the global magnetometer response \citep{dMC}, the ionospheric TEC disturbances and possible mechanisms driving them \citep{STR}, and the extreme auroral electrojet activity associated with supersubstorm processes \citep{Zou}. These studies demonstrate that the storm involved highly dynamic MI coupling processes and complex solar wind drivers rather than a single CME impact.

\begin{figure}
\centering
\noindent\includegraphics[width=0.9\textwidth,height=0.9\textwidth]{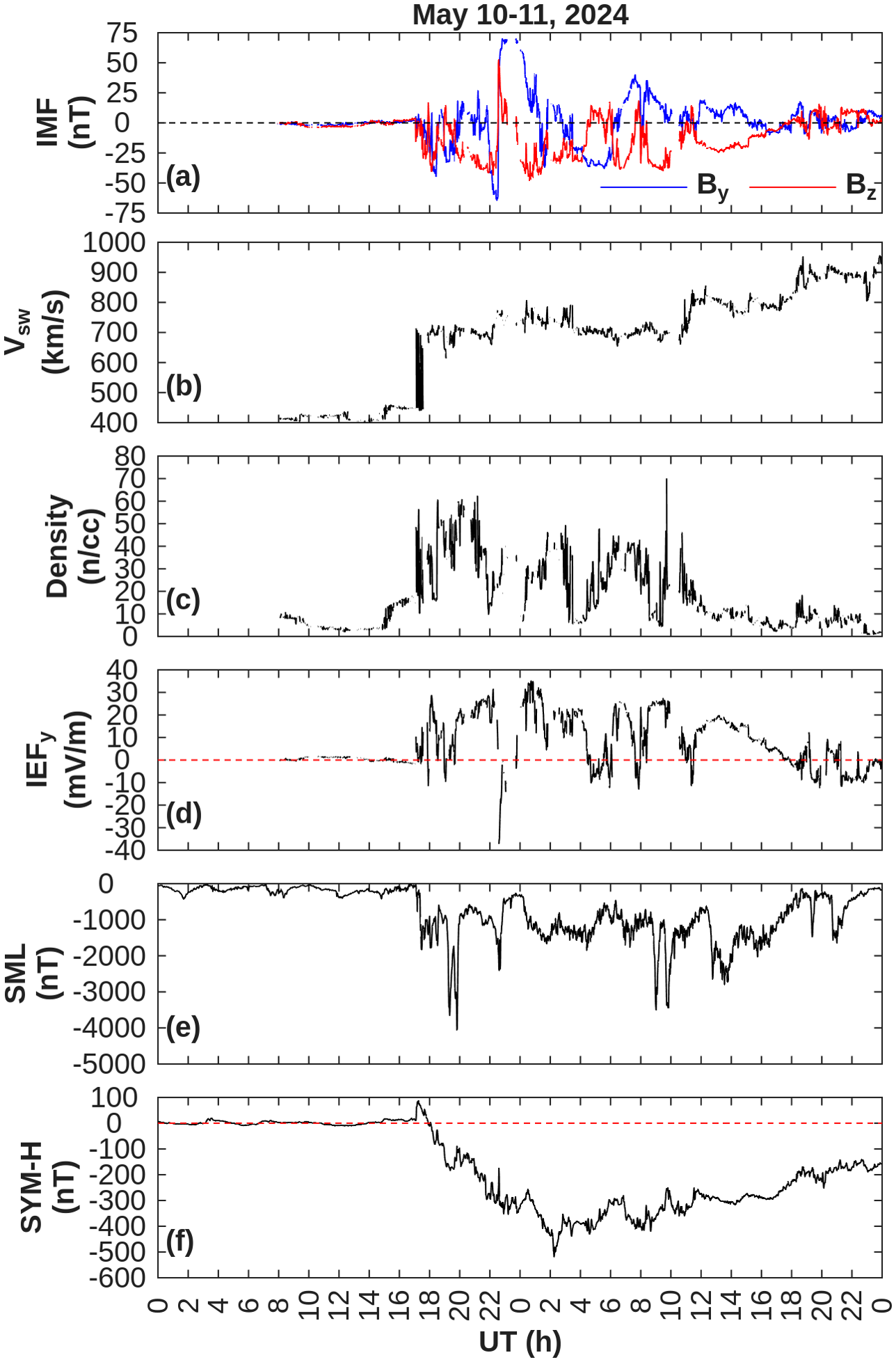}
\caption{Variations in: (a) IMFs (nT) $B_z$ (red) and $B_y$ (blue), (b)the $V_{sw}$ (km/s), (c) the density (n/cc), (d) the $IEF_y$ (mV/m), (e) the SML, and (f) the SYM-H (nT) during May 10-11, 2024.}
\label{m1}
\end{figure}

Figure \ref{m1} shows the interplanetary conditions together with the SML and SYM-H indices during May 10-11, 2024. The Storm Sudden Commencement (SSC) occurred as a result of the first ICME shock wave arriving at 17:05 UT at the magnetosphere. The compression of the dayside magnetopause currents \citep{Araki_T} produced a positive excursion in SYM-H (panel (f)), which reached up to 88 nT at 17:15 UT. The main phase began at 17:54 UT with a SYM-H value of -7 nT and continued until 02:14 UT on May 11, when SYM-H reached its minimum of -518 nT. The corresponding SML value was -1243 nT (panel (e)), indicating enhanced auroral electrojet activity. During this period, a combination of southward IMF, elevated solar wind speed, increased density, and large $IEF_y$ (panels (a) to (e) of Figure \ref{m1}) provided strong and continuous solar-wind driving, resulting in a pronounced ring-current intensification and enhanced ionospheric electrodynamic response.

While previous studies have described the global magnetospheric and ionospheric impacts of the May 2024 storm, fewer investigations have examined the simultaneous interhemispheric response of FACs, ground magnetic perturbations, and TEC variations using a combined multi-instrument approach. In the following analysis, we focus on this aspect by comparing AMPERE-derived FAC morphology with near-conjugate magnetometer and TEC observations from Antarctic and Arctic stations. This approach allows us to examine how storm-time electrodynamic disturbances manifest in both hemispheres.

\begin{figure*}
\centering
\noindent\includegraphics[width=300pt,height=500pt]{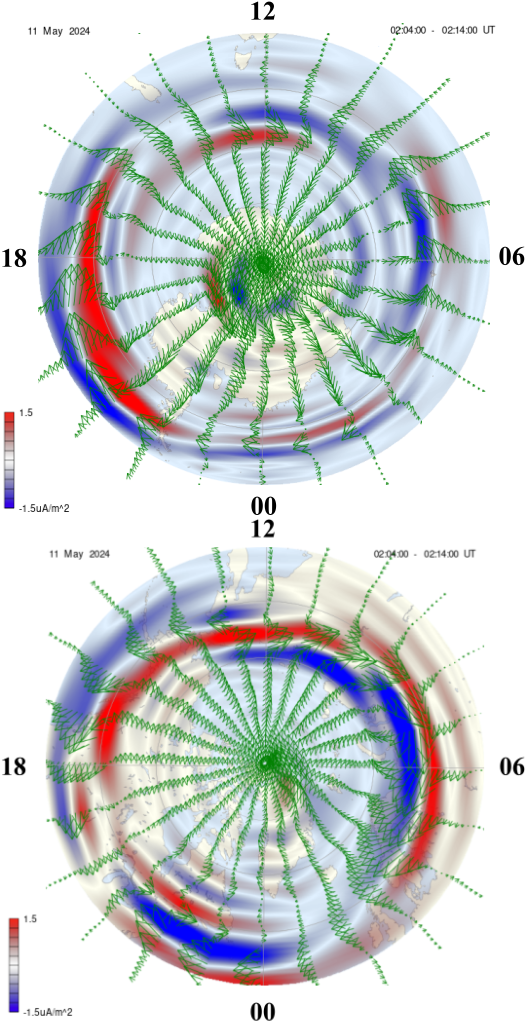}
\caption{AMPERE polar maps for the interval 02:04–02:14 UT on May 11, 2024, showing magnetic perturbation vectors (green) and the FAC distributions over the southern hemisphere (top panel) and northern hemisphere (bottom panel). Red indicates currents flowing out of the ionosphere, while blue indicates currents flowing into the ionosphere. The color bar represents the FAC density ($\mu A/m^2$). The maps provide a snapshot of the FAC morphology near the SYM-H minimum, illustrating the large-scale current configuration during this interval.}
\label{m2}
\end{figure*}

In Figure \ref{m2}, we present the AMPERE polar maps for the interval 02:04–02:14 UT on May 11, 2024, near the SYM-H minimum (02:14 UT), marking the end of the main phase of the Gannon storm. In these polar maps, the top and bottom correspond to noon (12 MLT) and midnight (00 MLT), respectively, while the left and right sides correspond to dusk (18 MLT) and dawn (06 MLT). The southern hemisphere map is mirrored in MLT to maintain a consistent dawn-dusk orientation with the northern hemisphere. The color scale represents the FAC density (ranging from -1.5 to +1.5 $\mu A/m^2$), where blue indicates currents flowing into the ionosphere and red indicates currents flowing out of the ionosphere. The FAC distribution exhibits the familiar Region 1 (R1) and Region 2 (R2) current configuration encircling the polar caps. In the southern hemisphere (top panel), the FAC system shows a broad region of enhanced currents spanning the dusk-to-post-midnight sector, together with additional current structures near the dawn sector. In the northern hemisphere (bottom panel), the FACs appear more spatially extensive, with prominent current structures on both the dawn and dusk flanks and multiple arcs spanning the dayside and nightside sectors. The magnetic perturbation vectors (green) are predominantly aligned tangentially around the polar cap, reflecting horizontal ionospheric current systems associated with auroral electrojets that close these FACs. Overall, the AMPERE maps illustrate the spatial organization of geomagnetic storm-time FAC systems around this period. Differences in the spatial distribution and intensity of the current structures between the two hemispheres reflect the complex and dynamic nature of MI coupling during geomagnetic storms. However, the maps shown here represent only a snapshot of the current morphology during the selected time interval.

\begin{figure*}
\noindent\includegraphics[width=\textwidth,height=150pt]{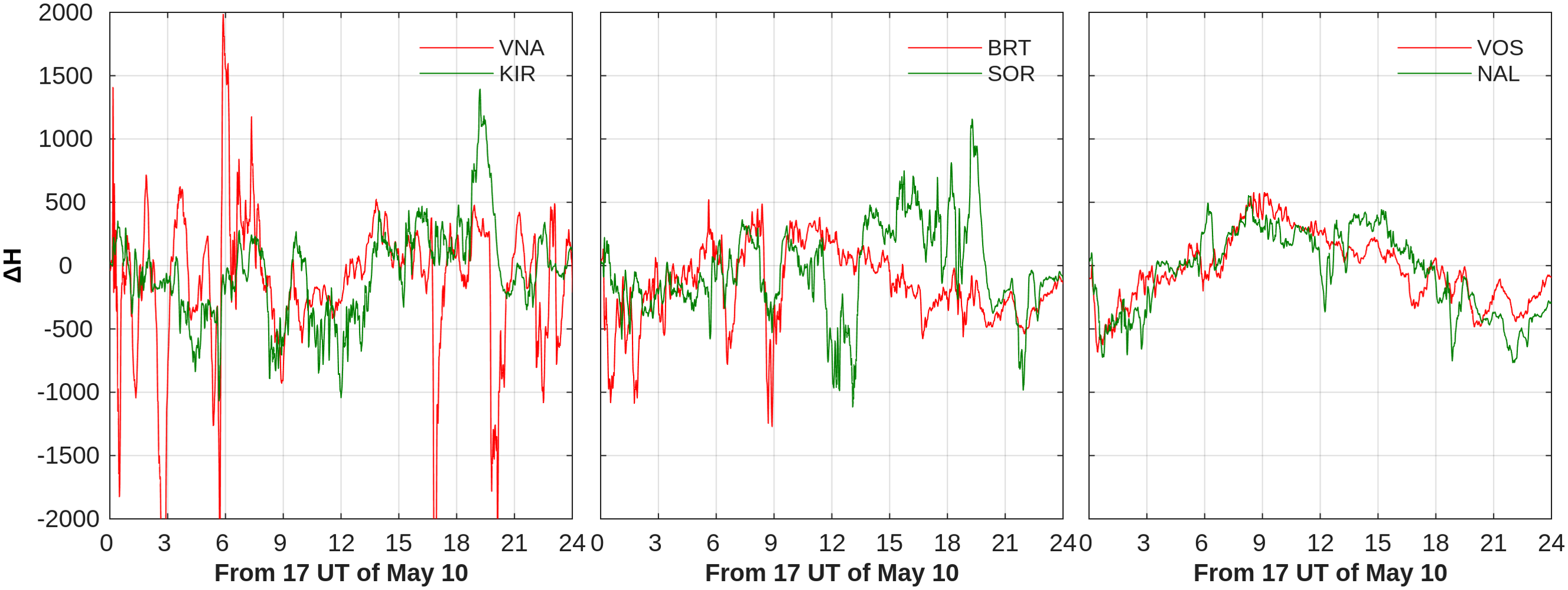}
\caption{$\Delta H$ variations over the Antarctic (red) and approximately magnetically conjugate Arctic (green) station pairs: VNA-KIR (left panel), BRT-SOR (middle panel), and VOS-NAL (right panel) during 17 UT of May 10 to 17 UT of May 11, 2024.}
\label{m3}
\end{figure*}

Further, Figure \ref{m3} presents the $\Delta H$ variations over a 24-hour interval, starting at 17:00 UT on May 10, 2024, at three Antarctic magnetometer stations (VNA, BRT, and VOS; in red) and their near-conjugate Arctic counterparts (KIR, SOR, and NAL; in green). These ground magnetic perturbations provide complementary information to the FAC patterns, as geomagnetic storm-time ionospheric currents and the associated FAC system produce corresponding magnetic signatures at high-latitude ground stations. It is to be noted that the stations are selected because they form near-conjugate pairs between Antarctica and the Arctic, enabling direct interhemispheric comparison of geomagnetic perturbations during the storm interval. The panels are arranged by longitude, moving from west to east Antarctica, left to right. This interval is selected because it includes the SSC, SYM-H minimum, and the early extended storm recovery phase. Each station pair reveals periods of clear hemispheric correspondence, simultaneous bays, and rapid fluctuations indicating a common magnetospheric driver. Yet, pronounced out-of-phase intervals are apparent, most clearly for VNA–KIR (left) and BRT–SOR (middle) pairs, where sharp perturbations in one hemisphere are delayed, weakened, or even opposite in sign compared to their conjugate. These phase differences indicate that the geomagnetic perturbations differ between the conjugate stations and vary with their geographic location and local time sector. Such differences may reflect hemispheric variations in ionospheric conductivity, local time conditions, or coupling to FACs, as suggested in previous studies \citep{sc:41,sc:42,Engebretson_MJ}. In contrast, the VOS–NAL (right) pair shows smoother variations but still features intermittent phase offsets. While the overall correspondence between conjugate station pairs reflects the influence of common large-scale magnetospheric driving, the observed differences highlight the role of regional ionospheric conditions in modulating the ground magnetic response. As a whole, Figure \ref{m3} demonstrates both conjugate symmetry and hemispheric asymmetry in geomagnetic responses from 17:00 UT on May 10 to 17:00 UT on May 11.

\begin{figure*}
\centering
\noindent\includegraphics[width=0.81\textwidth,height=0.81\textwidth]{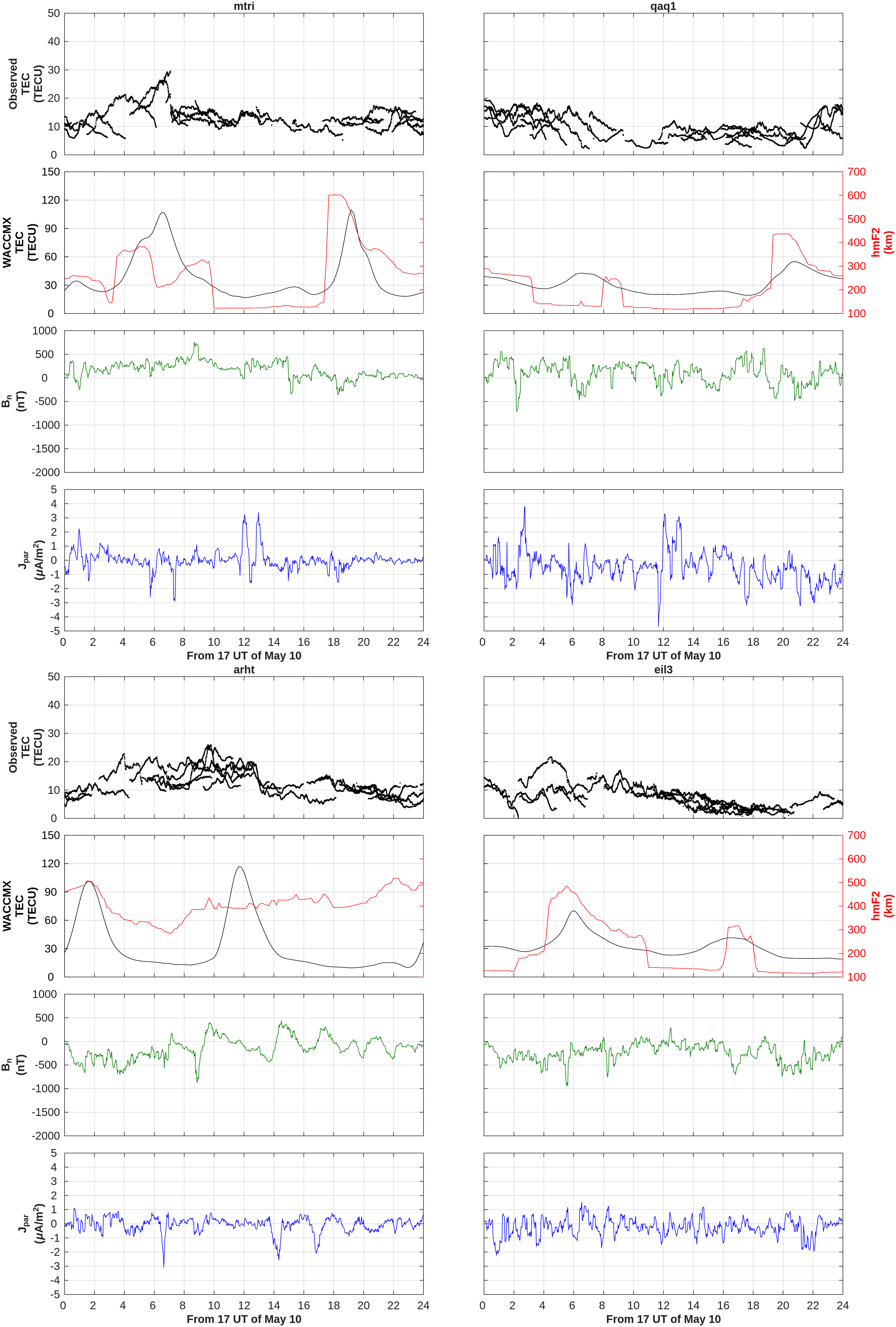}
\caption{The top and bottom panels correspond to the conjugate station pairs mtri-qaq1 and arht-eil3 from 17:00 UT (May 10) to 17:00 UT (May 11), 2024, illustrating the temporal evolution of ionospheric disturbances and associated electrodynamic activity at the high-latitude locations during the selected interval. Within the individual panels, the top row shows observed GPS TEC variations (black). The second row shows the WACCMX-simulated TEC (black) and hmF2 (red). The third and fourth rows show the AMPERE-derived $B_n$ (green) and $J_{par}$ (blue).}
\label{m4}
\end{figure*}

Finally, in Figure \ref{m4}, we present the observed GPS TEC variations at two Antarctic stations: 'mtri' and 'arht', and their near-conjugate Arctic counterparts: 'qaq1' and 'eil3'. These stations are selected based on the availability of continuous data for the period of interest. Rather than representing hemispheric averages, these plots illustrate the temporal evolution of TEC at the conjugate station pairs, allowing comparison of ionospheric responses between Antarctic and Arctic regions during the storm interval. The second panel shows the WACCMX-simulated TEC (black) and hmF2 (red), while the third and fourth panels display AMPERE observables: $B_n$ (green) and $J_{par}$ (blue). When comparing the pair mtri-qaq1, the station 'mtri' shows an increase in the TEC value around 00:00 UT on May 11. However, around this time, 'qaq1' shows no such enhancements. Moving to the other pair (arht-eil3), increases in TEC over 'eil3' can be observed around 21:00 UT, whereas TEC shows a maximum value at around 03:00 UT (May 11) over station 'arht'. The WACCMX simulations capture increases in TEC around the time when we observe enhancement over 'mtri' at around midnight UT. Similarly, over the station 'eil3', slightly delayed higher values of WACCMX TEC and corresponding hmF2 are observed, in comparison to the observed TEC rise at this station. The AMPERE-derived perturbations further illustrate the electrodynamic environment during the storm. The $B_n$ variations do not show significant variations, but the $J_{par}$ variations exhibit alternating positive and negative excursions, which are highest at around 05:00 UT on May 11 over 'qaq1', reflecting the evolving FAC systems during the disturbed period. Strong TEC disturbances and ionospheric variability during the May 2024 geomagnetic storm have been reported in previous studies \citep{dMC,STR}. The conjugate station comparisons presented here complement those studies with an observational perspective on how these storm-time disturbances manifested at the high latitudes over both hemispheres.

\newpage
\subsection{Severe event of October 10-11, 2024}

Following the Gannon event of May, in exactly five months, the second strongest storm of the year occurred during October 10-11. A high velocity CME (averaging around 1200-1300 km/s), associated with an X-class flare from the AR 3848 on October 9, was the primary driver of this geomagnetic storm when it reached the Earth at 15:20 UT on October 10, 2024. 

This particular storm has also been reported in recent studies examining the strong magnetospheric and ionospheric disturbances during the period of enhanced solar activity. Observations from satellite and ground-based measurements indicate that the storm produced significant ionospheric and thermospheric restructuring, including strong TEC variability. thermospheric drag and F-region uplift across multiple regions \citep{Denny,Paul,Tripathi_SC}. Studies of ionospheric current systems and electrojets during the event also indicate substantial changes in the high-latitude electrodynamic environment during the main phase of the storm \citep{Kleimenova}.

\begin{figure*}
\noindent\includegraphics[width=0.9\textwidth,height=0.9\textwidth]{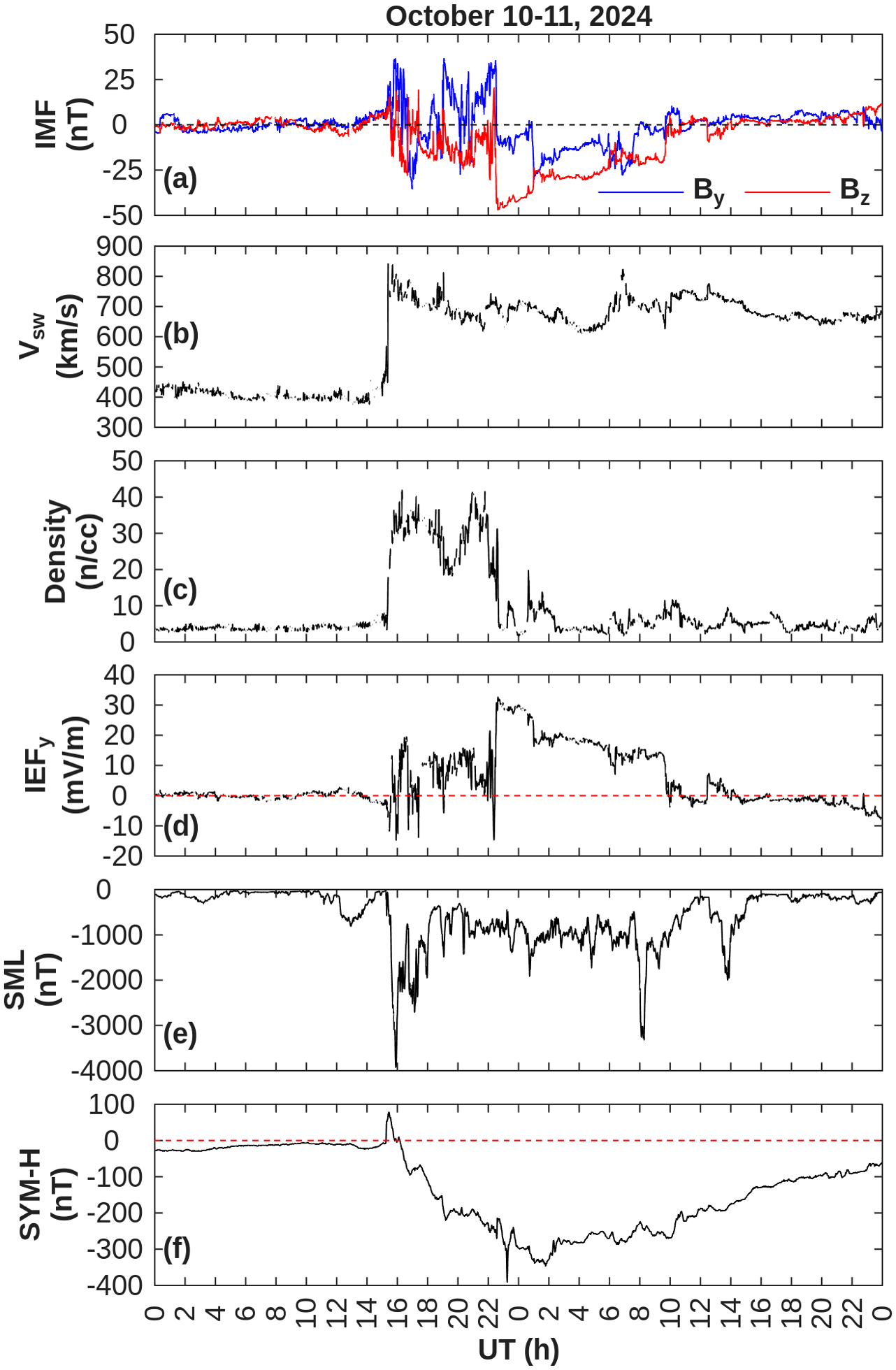}
\caption{Same as Figure \ref{m1} but during October 10-11, 2024.}
\label{o1}
\end{figure*}

Figure \ref{o1} shows the variations of the SML and the SYM-H indices along with the interplanetary conditions during October 10-11, 2024. The SSC occurred at 15:26 UT on October 10, marked by a sharp increase in SYM-H to 78 nT (panel (f)). Following the SSC, the geomagnetic storm main phase commenced once SYM-H reached -2 nT at 16:10 UT. SYM-H subsequently decreased rapidly and attained an initial minimum of –390 nT at 23:14 UT. The SML (panel (e)) simultaneously reached –577 nT, indicating enhanced auroral electrojet activity coincident with the peak ring-current intensification. However, this SYM-H minimum did not signify the termination of the main phase. Instead, SYM-H briefly recovered to –240 nT by 23:38 UT before declining again to a second minimum of –346 nT at 01:46 UT on October 11, marking the end of the main phase for this event. The stronger (more negative) values of SML compared to the earlier SYM-H minimum suggest intensified ionospheric current systems, consistent with sustained strong driving and continued energy input into the MI system despite partial ring current recovery. 

\begin{figure*}
\centering
\noindent\includegraphics[width=300pt,height=500pt]{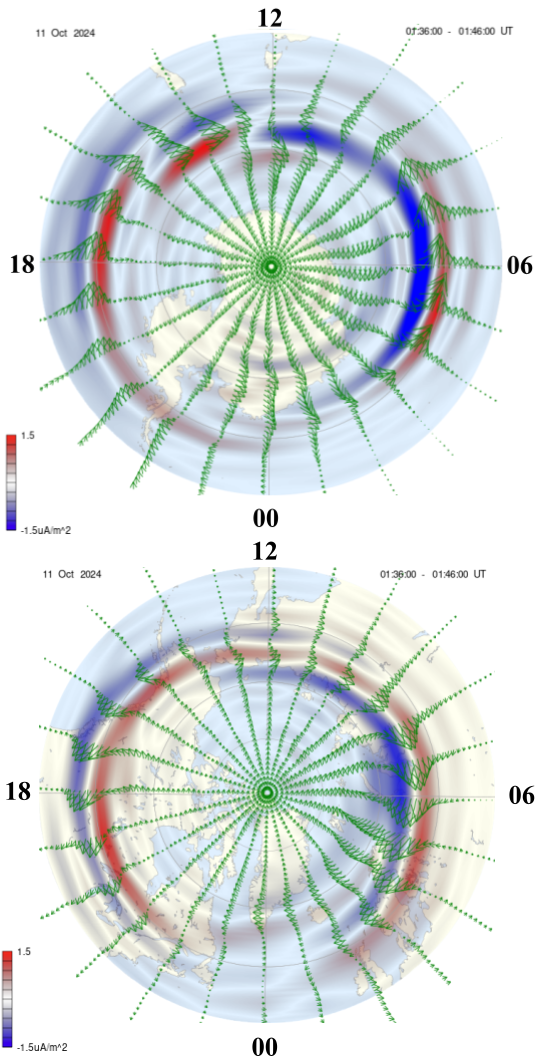}
\caption{Same as Figure \ref{m2} but for the interval 01:36-01:46 UT on October 11, 2024.}
\label{o2}
\end{figure*}

Figure \ref{o2} shows the AMPERE polar maps for the interval 01:36-01:46 UT on October 11, near the SYM-H minimum at 01:46 UT. Similar to Figure \ref{m3}, the top panel represents the southern polar region and the bottom panel represents the northern counterpart. Compared with the May 2024 event presented earlier, the FAC structures during this interval appear more localized and unevenly distributed across MLT sectors. In the southern hemisphere (top panel), enhanced FAC activity is concentrated primarily in the dawn sector, forming a pronounced current arc near 06:00 MLT, with additional weaker structures present on the dusk side. In contrast, the northern hemisphere (bottom panel) shows stronger currents on the dawn side, where a pronounced R1/R2 current pair is evident, accompanied by broader and weaker current sheets on the dusk side. The magnetic perturbation vectors (green) are predominantly oriented tangentially around the polar cap, indicating closure for the R1 and R2 FACs provided by the horizontal ionospheric current systems. Overall, the contrasting FAC distributions between the two hemispheres indicate substantial variability in the high-latitude electrodynamics during this storm interval.

\begin{figure*}
\noindent\includegraphics[width=\textwidth,height=150pt]{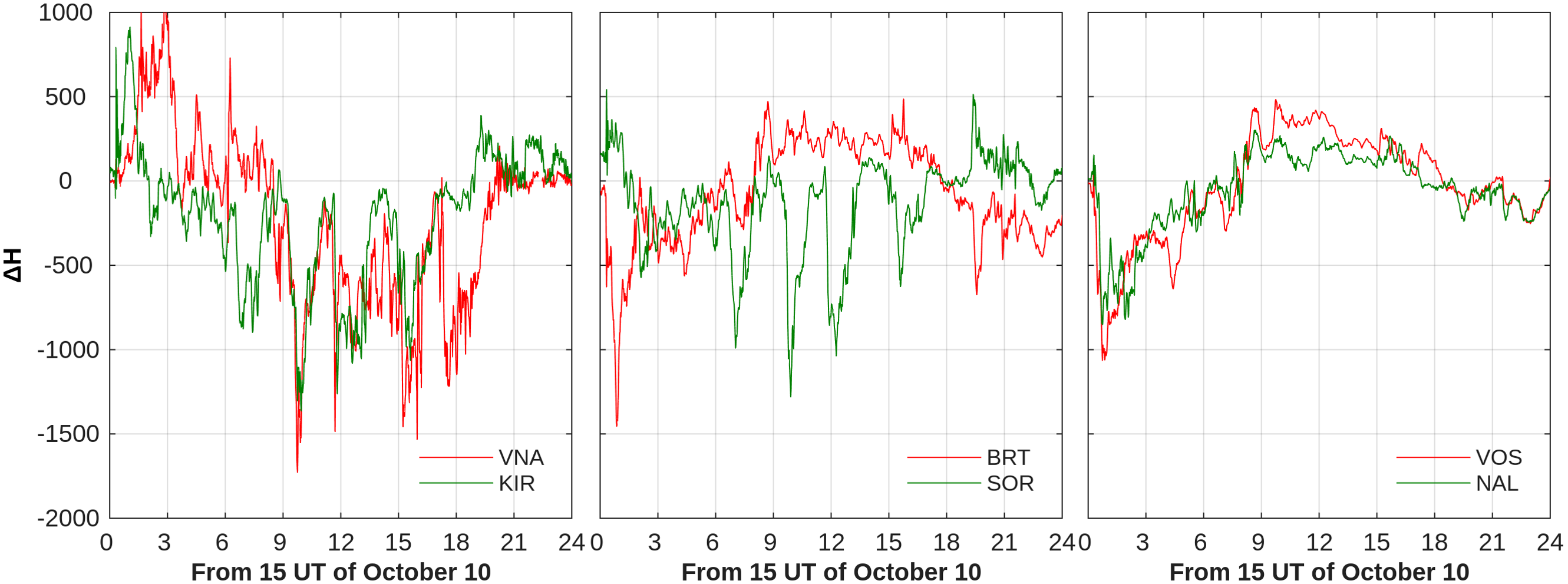}
\caption{Same as Figure \ref{m3} but during 15 UT on October 10 to 15 UT on October 11, 2024.}
\label{o3}
\end{figure*}

Figure \ref{o3} shows the $\Delta H$ perturbations at the three Antarctic magnetometer stations (VNA, BRT, and VOS shown in red) together with their near-conjugate counterparts in the Arctic region (KIR, SOR, and NAL shown in green) over a day, beginning at 15:00 UT on October 10, 2024. The panel layout and time interval follow the same convention used for Figure \ref{m3}. The figure shows broadly similar storm-time responses at the conjugate station pairs, although notable hemispheric and station-specific differences are also evident. Across all station pairs, enhanced geomagnetic activity begins shortly after 15:00–15:30 UT on October 10, corresponding to the onset of increased geomagnetic disturbance. However, both the amplitude and temporal evolution of the $\Delta H$ variations differ between the hemispheres. For the VNA–KIR pair (left panel), a pronounced negative perturbation is observed around 00:00-03:00 UT on October 11 (approximately 9–12 h after 15 UT). The BRT–SOR (middle panel) also shows storm-time depressions, although the Arctic stations exhibit larger negative excursions than their Antarctic counterparts. In contrast, the VOS–NAL pair (right panel) shows comparatively smoother variations throughout the interval. These observations indicate that while the conjugate stations often respond to the same storm-time disturbance, the amplitude and detailed temporal structure of the perturbations can vary between hemispheres. These observations indicate that while the conjugate stations often respond to the same storm-time disturbance, the amplitude and detailed temporal structure of the perturbations can vary between hemispheres. Similar hemispheric differences in conjugate magnetometer responses have been reported in previous studies. They are often discussed in terms of factors such as local time conditions, ionospheric conductivity differences, and variations in electrodynamic coupling. However, such mechanisms cannot be uniquely determined from the present observations alone. To further examine the ionospheric response during the October storm, we next analyze the temporal evolution of TEC at approximately magnetic conjugate station pairs.

\begin{figure*}
\centering
\noindent\includegraphics[width=0.81\textwidth,height=0.81\textwidth]{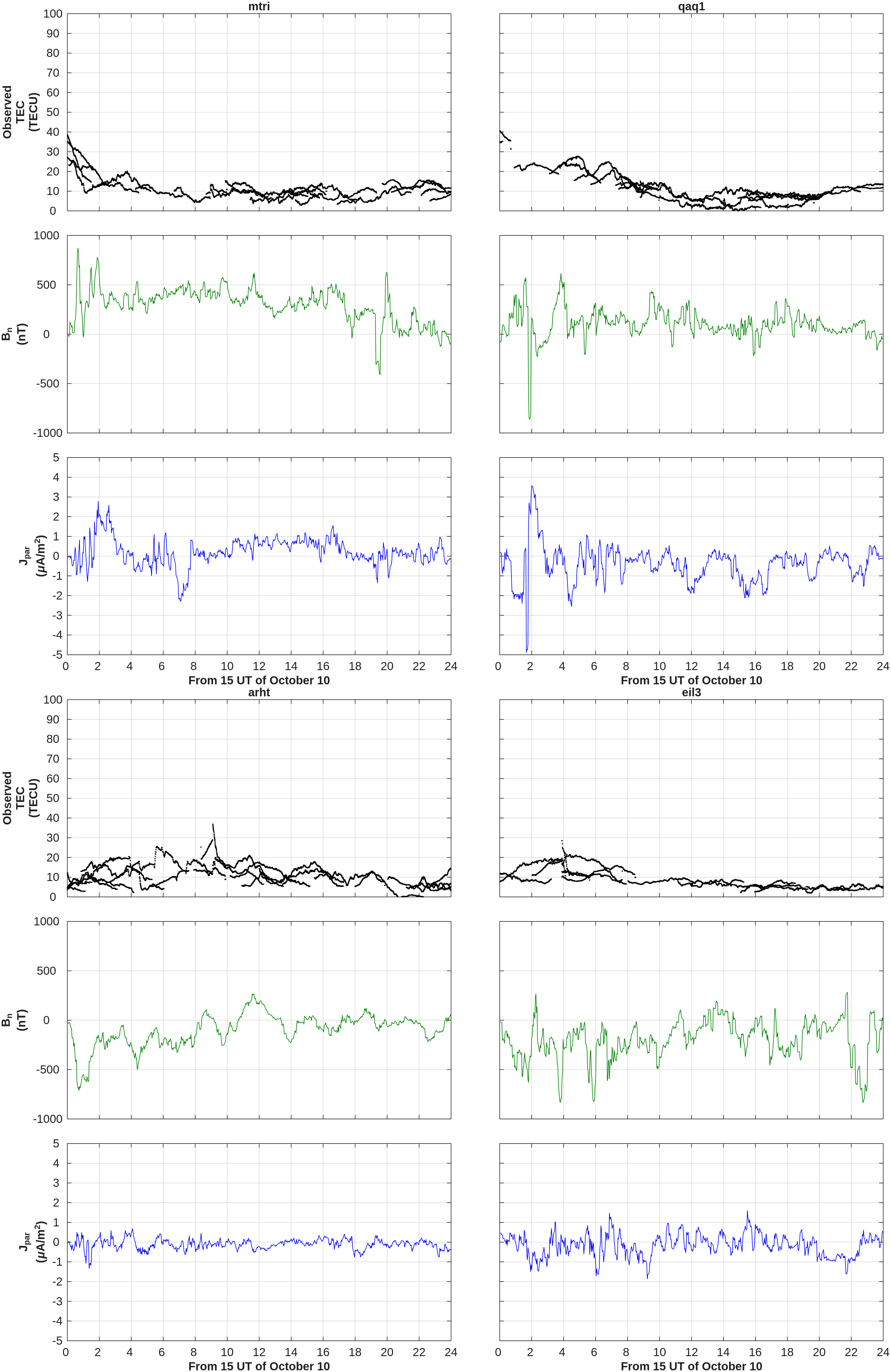}
\caption{Same as Figure \ref{m4} without the WACCMX-simulated variations and during 15 UT on October 10 to 15 UT on October 11, 2024.}
\label{o4}
\end{figure*}

Finally, Figure \ref{o4} shows the observed GPS TEC variations (top panels, black), together with the AMPERE-derived $B_n$ (green) and $J_{par}$ (red) in the middle and lower panels, respectively, for the same station pairs shown earlier in Figure \ref{m4}. The 24-hour interval begins at 15:00 UT on October 10, 2024. The WACCMX simulations of TEC and hmF2 cannot be shown because model runs are unavailable for this interval. The mtri–qaq1 pair shows lower TEC levels and a gradual decline, exhibiting steady variations. For the arht–eil3 pair, the fluctuations appear slightly sharper at 'arht', with an enhancement at around 01:00 UT on October 11, while the TEC variation at 'eil3' is flat. Looking into the magnetic perturbations, $B_n$ (in green), we see a sharp negative excursion up to -800 nT at 17:00 UT on October 10 over 'qaq1'. At around the same time, a negative followed by a positive transition (-6 to +3 $\mu A/m^2$) can be seen in the current density, $J_{par}$ (blue), over this station. Fluctuations in $B_n$ can be seen over all four stations, whereas $J_{par}$ is almost flat over 'arht', unlike its counterpart 'eil3'. Overall, these subplots indicate that while all stations recorded ionospheric disturbances during the storm interval, the magnitude and temporal evolution of TEC, magnetic perturbations, and FAC signatures varied between the station pairs. Such differences highlight the spatial variability of the ionospheric response and the complex electrodynamic conditions during the disturbed period from 15:00 UT on October 10 to 15:00 UT on 11 October 11, 2024.

\newpage
\section{Discussion}

The two geomagnetic events (May 10-11 and October 10-11, 2024) examined in this study provided a rare opportunity to investigate MI coupling processes under unusually intense and sustained solar wind driving. Both storms reached SYM-H values well below –300 nT, designating them among the strongest geomagnetic disturbances in decades. The observations presented here illustrate different manifestations of the MI coupling chain during the two storms. The AMPERE maps characterize the large-scale FAC morphology that transfers magnetospheric stress to the high-latitude ionosphere. These FAC systems close through horizontal ionospheric currents, which produce the ground magnetic perturbations observed at the magnetometer stations. The associated electrodynamic forcing modifies ionospheric plasma transport through enhanced convection and Joule heating, leading to spatially variable TEC responses. Therefore, differences in FAC morphology between the two storms can influence both the magnetic perturbation patterns and the ionospheric TEC variability observed at the conjugate stations.

The AMPERE polar maps shown provide snapshots of the FAC morphology during selected 10-minute intervals near the SYM-H minimum, when the MI coupling and associated current systems are typically most intense. While these maps do not capture the full temporal evolution of the current systems during the storms, they illustrate the spatial organization of the R1/R2 current system commonly observed during periods of strong geomagnetic activity. Such configurations reflect large-scale coupling between the magnetosphere and high-latitude ionosphere under enhanced solar wind driving \citep{Anderson_BJ,Anderson_BJ1}.

For the May event, the FAC pattern appears relatively broad and organized, with extensive R1/R2 current ovals surrounding the polar caps and strong current arcs across both dawn and dusk sectors as well as the dayside and nightside regions. Such large-scale current systems are consistent with enhanced storm-time FAC activity that commonly occurs during periods of strong dayside reconnection and ring current development, as reported in statistical AMPERE analyses \citep{Coxon_JC}. In contrast, the October event shows a more localized and asymmetric FAC structure. In the southern hemisphere, a relatively narrow current arc appears concentrated in the dawn sector, whereas in the northern hemisphere, a pronounced dawn-side R1/R2 pair is visible together with weaker and broader current sheets extending toward the dusk sector. These uneven structures suggest the presence of mesoscale FAC variability during this interval. Previous AMPERE studies have shown that storm-time FAC density distributions often exhibit heavy-tailed behavior \citep{Coxon_JC2}, meaning that localized current enhancements occur more frequently than expected for a normally distributed current system. The October map displays notable north-south and dawn-dusk asymmetries, with FAC activity concentrated near the dawn sector in both hemispheres but with a stronger and more structured R1/R2 pair in the northern hemisphere and a more localized arc in the southern hemisphere. This contrasts with the more spatially extensive and relatively organized FAC patterns observed during the May event. Such asymmetries may reflect several factors, including the influence of non-zero IMF $B_y$, differences in ionospheric conductance associated with seasonal illumination, and variations in magnetospheric stress redistribution that can lead to hemispheric and local time asymmetries in FAC and convection patterns \citep{Anderson_BJ,Anderson_BJ1,sc:42,sc:55}. Because FAC systems evolve significantly during geomagnetic storms, the spatial patterns shown here should be interpreted as illustrative examples of geomagnetic storm-time current morphology rather than a complete representation of the interhemispheric response.

Next, upon examination of the conjugate $\Delta H$ responses to the two events shown, geomagnetic disturbances occur nearly simultaneously at the conjugate station pairs. For the May event, the three polar conjugate pairs exhibited intervals of hemispheric correspondence, including broadly similar bays and rapid fluctuations, consistent with common magnetospheric driving (i.e., the global response of the magnetosphere to enhanced solar wind and IMF forcing during geomagnetic storms). Such similarities in high-latitude magnetic perturbations at conjugate locations have been reported in previous studies of interhemispheric geomagnetic responses, including Pc5 wave observations whose ground signatures often appear simultaneously in both hemispheres \citep{Pilipenko_VA}. For the October storm, the general pattern of nearly simultaneous responses is again visible, particularly following the interplanetary shock impact. All four station pairs show enhanced activity shortly after the SSC, consistent with a global magnetospheric response during the storm main phase. However, the amplitude and temporal evolution of the $\Delta H$ perturbations vary considerably among the station pairs. These differences indicate that although conjugate stations often respond to the same storm-time disturbance, the detailed magnetic signatures can vary between hemispheres and locations. Previous studies suggest that such variations may be influenced by several factors, including differences in ionospheric conductivity associated with seasonal illumination (i.e., hemispheric differences in solar illumination that affect ionospheric conductivity and therefore the strength of ionospheric current systems), variations in local electrodynamic geometry (i.e., differences in magnetic latitude, local time sector, and geomagnetic field configuration that influence the mapping of ionospheric currents to ground magnetic perturbations), and asymmetries in FAC closure (i.e., differences in how FACs couple to ionospheric current systems in the two hemispheres due to conductivity or magnetic field differences) \citep{Workayehu_AB,Engebretson_MJ,sc:41,sc:42}. In this context, the observed station-to-station differences likely reflect the complex and spatially variable electrodynamic conditions that develop in the high-latitude ionosphere during geomagnetic storms.

Furthermore, as ionospheric current systems and FACs are closely coupled with high-latitude plasma transport and electrodynamic forcing, the magnetic perturbations observed at the conjugate stations are often accompanied by corresponding changes in the electron density. We therefore examined the TEC responses at the conjugate station pairs to further investigate how the ionospheric plasma distribution evolved during these storm intervals. For the May event, the conjugate response reflects a distinct time-offset and spatially asymmetric evolution of high-latitude electrodynamics, where enhancements are not simultaneously mapped between hemispheres but instead occur at different UT sectors across conjugate pairs. This behavior is consistent with the passage of structured auroral precipitation and convection features through the polar cap, producing localized F-region uplift and density enhancements that are further modulated by background conductivity differences. The partial agreement with WACCMX, capturing some timing but with delayed responses at specific locations, supports the interpretation that the dominant drivers are transient and regionally evolving rather than globally coherent. The AMPERE signatures reinforce this view of the absence of strong, systematic $B_n$ perturbations alongside pronounced, alternating $J_{par}$ excursions, indicating dynamically evolving FAC systems with significant temporal restructuring, characteristic of active R1/R2 current systems during disturbed intervals. In contrast, the October event exhibits a weaker and more unevenly distributed response, where reduced TEC levels and smoother evolution in some conjugate pairs coexist with sharper, localized perturbations in others, pointing to less efficient interhemispheric coupling. The occurrence of isolated but strong magnetic deflections, together with spatially inconsistent $J_{par}$ behavior, including near-flat responses at certain locations, suggests that current systems are more fragmented and locally controlled. Altogether, these differences highlight that polar conjugacy during storms is governed by the timing and morphology of auroral forcing and FAC evolution, leading to event-specific asymmetries rather than uniform hemispheric responses.

The contrast between the two events is broadly consistent with previous studies indicating that storm-time ionospheric responses can vary significantly between hemispheres and longitudes, even under strong magnetospheric driving. Such variability may be influenced by differences in FAC distributions, ionospheric conductance, neutral wind circulation, and geomagnetic field geometry, which numerical and observational studies have identified as important factors contributing to high-latitude ionospheric asymmetries \citep{Hong_Y}. Taken together, the May and October storms illustrate that while both events imposed strong, large-scale magnetospheric forcing, the resulting TEC response ranged from relatively coherent behavior during the May event to more heterogeneous and localized responses during the October storm. These differences highlight the importance of local ionospheric and geomagnetic conditions in shaping regional variations in MI coupling.

As a path forward, this work can be extended through multi-event statistical analyses aimed at isolating the roles of ionospheric conductance, IMF orientation, and mesoscale FAC variability in shaping conjugate responses during periods of strong geomagnetic activity. The incorporation of data-assimilative frameworks that integrate AMPERE observations, ground magnetometer measurements, and GPS-derived TEC with physics-based ionosphere–thermosphere coupled models could further improve our understanding of how localized current structures evolve and interact with thermospheric composition and neutral wind dynamics. Applying such approaches across a broader set of geomagnetic storms, spanning different seasons and IMF conditions, would help identify systematic patterns in hemispheric asymmetry and contribute to improving the predictive capability for storm-time FAC distributions, TEC variability, and associated magnetic perturbations.

\section{Summary}

The geomagnetic storms of May and October 2024 provided an opportunity to examine MI coupling under unusually intense and sustained solar wind driving, revealing that similar large-scale forcing can produce markedly different ionospheric responses. AMPERE observations showed that both events were associated with enhanced R1/R2 FAC systems, but with contrasting spatial organization. The May event exhibited relatively broad, structured, and spatially extensive current systems, whereas the October storm displayed more localized, asymmetric, and uneven FAC morphology with pronounced hemispheric and dawn-dusk asymmetries, indicative of mesoscale variability. Ground magnetometer measurements indicated nearly simultaneous onsets of geomagnetic activity in both events, consistent with a global magnetospheric response. However, the amplitude and temporal evolution of the $\Delta H$ perturbations varied substantially between station pairs, reflecting the influence of local ionospheric conductivity, electrodynamic geometry, and asymmetric FAC closure. The TEC responses reinforced this contrast. The May event showed relatively coherent behavior with clear temporal offsets between conjugate locations, consistent with the passage of structured auroral precipitation and dynamically evolving convection and FAC systems, whereas the October event exhibited weaker, more heterogeneous, and spatially inconsistent TEC variability, suggesting less efficient and more locally controlled interhemispheric coupling. Overall, these results demonstrated that storm-time MI coupling is not uniformly expressed across hemispheres even under comparable external driving, but is strongly modulated by the spatial morphology and temporal restructuring of FAC systems together with local ionospheric conditions, leading to event-specific differences ranging from organized large-scale responses to fragmented and asymmetric electrodynamic behavior.

\section*{Acknowledgments}

SC acknowledges the Department of Science and Technology (DST), Government of India, for providing a research fellowship. This work is part of the POlar and space WEather Research (POWER) project of the IIG and is supported by DST, India. The authors acknowledge the NASA OMNIWeb Data Explorer for the openly available high-resolution (1-min) IMF $B_z$, $B_y$, $V_{sw}$, density, $IEF_y$, and the SYM-H data. The authors would also like to thank the IIG and NCPOR, MoES staff for their logistical and scientific support in maintaining the database at the two Indian Antarctic stations (Bharati and Maitri). Acknowledgments are due to the SuperMAG network for providing the openly available SML index and the $\Delta H$ data. The authors thank the PIs of the magnetic observatories and the national institutes that support the observatories. The authors thank the IGS network for providing GPS TEC data openly. The NCAR WACCM-X model hosted by the CCMC is duly acknowledged for the simulation run during the May event. We thank the AMPERE team and the AMPERE Science Data Center for providing data products derived from the Iridium Communications constellation, supported by the National Science Foundation.

\bibliographystyle{jasr-model5-names}
\biboptions{authoryear}
\bibliography{SC.bib}

\end{document}